\documentclass{article}
\usepackage{graphicx}
\usepackage{epsfig}
\usepackage[usenames]{color}
\usepackage{amsmath, amssymb}
\usepackage{fancyvrb}
\usepackage{cite}
\usepackage{url}

\newcommand{\rr}{{\mathbb R}}

\DeclareFontShape{OT1}{cmtt}{bx}{n}{<5><6><7><8><9><10><10.95><12><14.4><17.28><20.74><24.88>cmttb10}{}

\newcounter{figctr}
\newcommand{\ilab}[1]{\immediate\write1{\string \newlabel{#1}{{\arabic{figctr}}{\thepage}}}}

\newcommand{\chhobi}[4]{
\parbox{\textwidth}{\begin{center}
  \rotatebox{#3}{\scalebox{#2}{\includegraphics{#1}}}\\
\underline{{\bf Fig \arabic{figctr}:\ \ilab{im:#1}\addtocounter{figctr}{1}}
  {\bf #4}}
\end{center}}
}

\title{RESID: A Practical Stochastic Model for Software Reliability}

\author{Arnab Chakraborty\\
{\em Applied Statistics Unit}\\
{\em Indian Statistical Institute}\\
{\em 203, B T Road}\\
{\em Kolkata, India 700108}\\
{\em Phone: 91-33-2334-0337}\\
{\tt arnabc@isical.ac.in}}

\begin{document}
\date{}
\maketitle

\abstract{
A new approach called RESID is proposed in this paper for
estimating reliability of a  software allowing for
imperfect debugging. Unlike
earlier approaches based on counting number of bugs or modelling
inter-failure time gaps, RESID focuses on the probability of
``bugginess'' of different parts of a program buggy. This perspective allows an easy
way to incorporate the structure of the software under test, as
well as imperfect debugging. One main design objective behind
RESID is ease of implementation in practical scenarios.  

\section{Introduction}
With computer programs pervading all walks of modern life,
software debugging has long been an area of active
interest. There are three major aspects to this problem. Firstly,
one needs better software development tools to avoid creating
bugs. Secondly, one needs to be able to detect and correct bugs
that have already crept in. The third goal is to estimate the
reliability of a software program. Since no fool proof debugging
method is known to exist, the third goal is no less important
than the first two. Various approaches have been suggested in the
literature to estimate the reliability of a given piece of
software, ranging from simple profiling techniques (disregarding
the stochastic nature of bugs) to
elaborate stochastic models (that often overlooks the structure
of the program). In this paper we propose a new technique
called {\bf Reliability Estimation for Software under Imperfect
  Debugging (RESID)} to make
the twain meet: a statistical method based on maximum likelihood
estimation that also takes the structure of the program into
account. The model also allows the possibility of imperfect
debugging, where a particular chunk of code is allowed to contain
bugs (albeit with a reduced probability) even after multiple debugging sessions.

The paper is laid out as follows. In the next section we review
various techniques proposed for the problem, with a brief
discussion of their merits and demerits. Section 3 presents the
new method from the theoretical viewpoint. Suggestions for
practical implementation of the method are given in section
4. Section 5 presents some discussion about
the performance of the technique based on simulation.
Section 6 deals with some variations of {\bf RESID} to suit specific
needs. After a brief concluding section some probabilistic
underpinnings of the method 
are outlined in an appendix.
 
\section{Review of existing techniques}
Many models and approaches have been suggested in the literature
to assess software reliability. We shall briefly a review a
selection of these techniques, without aspiring for
comprehensiveness, which anyway is beyond the scope of this short
paper. An extensive literature review is given in
\cite{xie}. 
    Software reliability is typically defined as the probability
of failure-free operation of a computer programme in a specified
environment for a specified period of time \cite{musabook}.
As \cite{bolandshort} points out the statistical
models for software reliability come in two distinct flavours,
those that deal with time between successive failures, and those
that deal with counting bugs in a program.

The former approach, pioneered by \cite{jelinski,moranda}, makes
the
(somewhat unrealistic) assumption that inter-failure times are
exponential in nature, and are independent of one
another. Methods of this genre also make the assumption that a
bug is always rectified when it is detected. This unfortunately
leaves no place for wrong or incomplete fixes, a phenomenon that
ubiquitously plagues the software industry. A related approach
is taken by \cite{singsoyer,singpurwalla}, where the
authors try to fit time series models to the inter-failure
times. 

The second approach uses point processes to model occurrences of
bugs\cite{goel}. The paper \cite{nayak} is a typical example, where the author employs Poisson processes for this purpose. One notable
feature of this approach is that the same bug is allowed to recur. Indeed, the author suggests that after detection
a bug should either not be removed or at least have its place 
marked, so that every subsequent pass through that position may
be counted. This idea has been
extended in \cite{dewanji} where the presence of multiple bugs is
modelled as multiple Poisson processes running independently in
parallel. The main interest there lies in estimating the number of
processes.  

However, as noted in \cite{xie}, the plethora of proposed,
pedantic models
contrasts sadly with the paucity of practically implementable
ones. Due to resource constraints, debuggers and programmers 
often have to resort to {\em ad hoc} plans to yield quick
results, often  in response to the needs of some
irritated customer demanding a fix for a particular bug. Most of
the methods proposed in the literature are a bit too elaborate to
cope with such real life scenarios. 
Some practical method
based on a reasonable statistical model would be a useful
addition to a software engineer's repertoire.
In this paper we seek to propose such a method.

\section{RESID}
One of the trickiest aspect of quantifying software reliability
is to come up with a quantitative definition of a bug. Often a
single mistake in a software triggers multiple branches of a
system to fail. Techniques based on only the occurrences of
failures often count each failed branch as a separate bug, while
from the viewpoint of software management these should be
considered as a single bug. We circumvent this inherent ambiguity
in the definition of a bug by looking instead at the concept of
``bugginess'' as follows. 

We can consider a piece of code as a flowchart with branches and
loops. Control may flow down different branches depending on the
initial data. However, every program consists of
some {\em chunks} which is defined as a sequence of consecutive
instructions without any embedded branching in it. Thus, once
control enters a chunk it must either flow through the chunk
along a unique linear path, or must crash the program (possibly
because of a bug in some earlier chunk.  

We shall assume that each chunk has some probability $p$ of
being ``buggy''. More specifically, $p$ is the chance that
we encounter some bug in that chunk while running the software
with a random data. Thus, strictly speaking randomness enters not
only due to the inadvertent errors of the programmers, but also
due to the choice of the input data. We shall also assume that the event
that one chunk contains a bug is independent of another chunk
containing a bug. This is not as impractical as it may
sound, as bugs are born of inadvertent mistakes on part of the
programmers, and not as a result of the control structure
relating the chunks. Our assumption of independence 
does not preclude the possibility that a bug in one chunk may
wreak havoc in a subsequent chunk. Mathematically inclined
readers not content with this explanation may see the appendix
for a more rigorous presentation of this idea.

To assess the reliability of a piece of software we start with
the structure of the program in terms of the chunks. During the
debugging phase the program is run multiple times, each time with
independent initial data. For each run we record the following
information.

\begin{enumerate}

\item whether the run has terminated correctly or not,
\item if the run indicates the presence of a bug, then which chunk
  contains the bug,
\item which chunks have been executed and how many times.

\end{enumerate}

The first two pieces of information are available from any
standard debugging session. In order to collect the frequency of
execution of the chunks one has to embed a logging command
in  each chunk. Such an embedding can be
easily achieved automatically using software tools. We assume
that a bug is fixed (possibly imperfectly) once it is
detected. We shall
later point out two variants of the same approach to cope with
situations where a bug cannot be removed, or where multiple bugs
are identified in a single run. 

Notice that even though we feed independent initial values in the
program each time, the collected data are not iid in nature,
since the underlying software changes with each debugging. 

We stochastically model the software debugging mechanism as
follows. Initially each chunk is believed to be buggy with
probability $p.$ 

Every time a chunk is ``debugged'' we
assume that the probability of its still remaining buggy is
scaled down by a known factor $\alpha\in (0,1),$ which
measures the debugging inefficiency. Thus, after detection and correction of its
first $k$ bugs, a
chunk has probability $p\alpha^k$ of remaining buggy.
A value of $\alpha$ close to 0 implies efficient debugging,
while a value close to 1 implies the opposite.

We shall consider $p$ as a measure of unreliability (or,
rather, the lack thereof) of the over all software. Each chunk
gets its own unreliability score $p\alpha^k.$ 
It is not difficult to come up with a color coding scheme to
depict the unreliability scores of the different chunks
diagrammatically, {\em e.g.}, using a UML diagram. 

If $p$ is sufficiently small we might consider the
software as reliable enough. If, however, $p$ is large, then
we may like to focus our debugging efforts on the chunks
with higher unreliability scores.

We shall employ maximum likelihood estimation to
estimate $p.$ Writing down the likelihood function, however,
is a bit tricky here, as we have to take the structure of the
software into account. The procedure is best explained with a
simple example. 

Consider a simple program with control flow as shown
in {\rm Fig \ref{im:flow1.eps}}. It has 3 chunks each labelled with
an arrow and number. The control starts in chunk 1, then comes to
an {\tt if}-class, and branches out into chunk 2 or 3. 

\chhobi{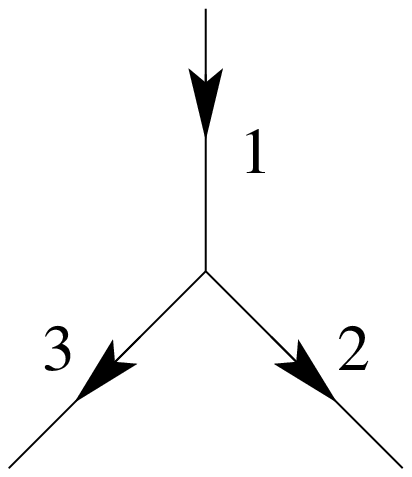}
       {0.4}
{0}
       {A 3-chunk
  program with an {\tt if}-clause}

The program is executed 5
times with the following results:

\begin{enumerate}

\item 1; bug in 1 ({\em i.e.}, the program crashed in chunk 1 due to a
  bug in that chunk)
\item 1,2; bug in 2
\item 1,3; no bug
\item 1, 2; bug in 1 ({\em i.e.}, the program continued up to chunk 2,
  but a bug was found in chunk 1)
\item 1,3; bug in 3

\end{enumerate}

\newcommand{\bug}{\text{ bug }}
\newcommand{\nobug}{\text{ no bug }}
We assume that the buggy chunk is (imperfectly) debugged after
each unsuccessful run. Also, the chunks visited after passing
through a buggy
chunk produce unreliable results, and so are to be ignored. For
example, we shall truncate the record for the fourth run above 
to 

\begin{quote}

1; bug in 1.

\end{quote}

The
probability of bugginess of each chunk is listed below, along
with the likelihood values for each run:

\begin{center}

\begin{tabular}{c|ccccccccccccccc}
\hline\hline

Stage & Chunk 1 & Chunk 2 & Chunk 3 & Likelihood\\
\hline
0 & $p$ & $p$ & $p$\\
1 & $p\alpha$ & $p$ & $p$ & $P(1; \bug)=p$\\
2 & $p\alpha$ & $p\alpha$ & $p$ & $P(1,2;\bug)=(1-p\alpha)p$\\
3 & $p\alpha$ & $p\alpha$ & $p$ & $P(1,3;\nobug)=(1-p\alpha)(1-p)$\\
4 & $p\alpha^2$ & $p\alpha$ & $p$ & $P(1;\bug) = p\alpha$\\
5 & $p\alpha^2$ & $p\alpha$ & $p\alpha$ & $P(1,3;\bug)=(1-p\alpha^2)p$
\\
\hline
\end{tabular}
\end{center}

Taking the product of the last column we get the likelihood of
the entire data set as
$$
L(p) = \text{constant}\times p^4(1-p)(1-p\alpha)^2(1-p\alpha^2).
$$
 It is not hard to see that this scheme can be generalized to any
 branching structure. However, the situation becomes somewhat
 different in presence of loops. A bug inside a loop may not be
 triggered during the very first pass through the loop. In practice it is often
 difficult to keep track of the exact pass when a bug inside a loop is
 triggered for the first time. So if the program halts due to a
 buggy chunk inside a loop,
 all that we can be sure of is that the chunks
 leading to it have worked correctly at least once. We have no
 way of knowing if the bug has already been triggered before
 subsequent passes through the loop or not. We explain this idea
 with an example. 

Consider the loop structure shown
in {\rm Fig \ref{im:loop1.eps}}. 

\chhobi{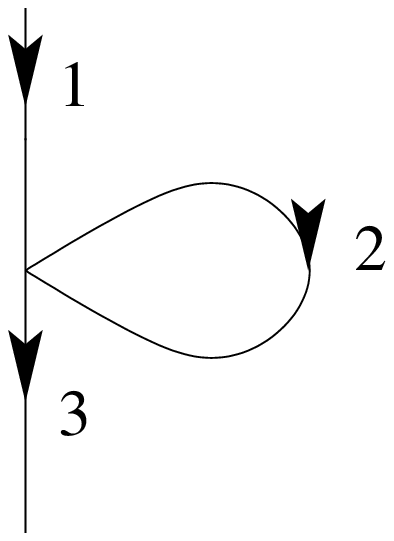}
       {0.4}
{0}
       {A 3-chunk prgroam
  with a
  loop}

Suppose that a run through this program produces the following
record:

\begin{quote}

1,2,2,2,3; bug in 2.

\end{quote}

As should be obvious to anybody with even
moderate debugging experience, it is hard to
detect which pass(es) through chunk 2 has (have) triggered the
bug. As a result all that we can be sure of is that chunk 1 has
worked fine in this run, with at least one pass of chunk 2
failing. Once a bug is triggered the subsequent chunks cannot be
reliably debugged. So we shall truncate the record to 

\begin{quote}

1,2; bug in 2.

\end{quote}

As can be seen easily the likelihood under this model is of the
form 
$$
L(p) \propto p^m \prod_{i=0}^k (1-p\alpha^i)^{n_i}.
$$

Here $k$ is the maximum number of debugging session for any
chunk, $m$ is the number of bugs detected and removed,
and $n_i$ (for $i=0,1,...)$ is the number of
perfect runs for chunks with exactly $i$
debugging attempts. 
 
The log-likelihood is (up to an additive constant) 
$$
\ell(p) = m\log p + \sum_{i=0}^k n_i\log(1-p\alpha^i).
$$

The following fact is useful for numerically maximising this
function for $p\in(0,1).$

\medskip

\noindent 
{\bf Lemma:}\  For any program and any debugging outcomes, the model
  has a strictly concave log-likelihood function. ///

\medskip

\noindent 
{\em Proof:}\ 
The functions $\log p$ and  $\log (1-\frac{p}{a})$ are 
strictly concave for any $a>0,$ and a linear combination of
concave functions with positive coefficients is again strictly
concave.
///

This fact implies the uniqueness of MLE $\hat{p},$ if it
exists. Unfortunately, MLE may not always exist. But it does
exist under the following fairly mild condition.

\noindent

\medskip

\noindent 
{\bf Lemma:}\ If $m,n_0>0$ 
then 
  $\ell(p)$ has unique maximum over $(0,1).$///

\medskip

\noindent 
{\em Proof:}\ 
This is because 
\begin{eqnarray*}
\lim_{p\to0+} \frac1p &=& \infty\\
\lim_{p\to0+} \frac{1}{1-p\alpha^i} &\in& \rr \text{ for } i=0,1,...
\end{eqnarray*}
Also
\begin{eqnarray*}
\lim_{p\to1-} \frac1p &\in& \rr\\
\lim_{p\to1-} \frac{1}{1-p} &>& \infty\\
\lim_{p\to1-} \frac{1}{1-p\alpha} &\in& \rr \text{ for } i=1,2,...
\end{eqnarray*}
So if $m,n_0>0,$ we have
$$
\ell'(0+) > 0 \text{ and }  \ell'(1-) < 0. 
$$
Strict concavity from the last lemma now clinches the argument.
///

The condition of this lemma has a simple
interpretation: $m>0$ means at least one bug is encountered
somewhere during some run of the program. The
condition $n_0>0$ means at lest one chunk has worked
properly in the very 
first attempt. It is easy to see that the probability of both
these happening goes to 1 as the number of chunks go to
infinity. 

Incidentally, it may be shown without much additional effort that
the condition $m>0$ is necessary for the existence of
MLE. However, the condition $n_0>0$ is only sufficient. In
fact, this is one of a general class of sufficient conditions of
the form $n_i > \alpha^{-i}-1.$ 

One
may now easily apply numerical methods like Newton-Raphson to
solve
$$
\ell'(p) = 0,
$$ 
or
$$
\frac{m}{p} - \sum_{i=0}^k \frac{n_i\alpha^i}{1-p\alpha^i} = 0.
$$
However, here we can avoid computing the second derivative needed for
Newton-Raphson iteration by using bisection method. 
 By virtue of the last lemma
 we can perform bisection method over the
interval $[\epsilon,1-\epsilon]$ for some suitably
small $\epsilon>0.$

\section{Implementation}
The main aim of this paper is to propose a systematic debugging
technique that can be easily integrated with existing
methods. Our exposition so far has been primarily theoretical. 
This section outlines how our
approach fits into a typical
debugging session, where 
input data may come from a customized design or user feedback
or simply generated randomly.

The first step is to identify the chunks in the software. This
can be achieved easily by simple lexical analyzers and parsers (like those generated
by {\bf flex} and {\bf bison}\cite{appel}). Such preprocessing steps are common in many
program validating scenarios. In our case the
preprocessing step creates a data base of the chunks, associating
each chunk with its file name and the first and last line
numbers. It also embeds a data logging command at the start of
the each chunk, such that the chunk number is recorded in a log
file the moment control enters that chunk. 

After this simple  preprocessing is over the actual debugging
starts, which consists of repeated runs each time with fresh
initial data.  After each run the programmer checks if the output is OK
or not. If it is, then this fact is recorded. If something has
gone wrong then the
programmer debugs the program, and identifies the line(s) that
required correction. Thus, for each run we get the logged record
of visited chunks, as well as the location of the bug, if any.

The resulting data set is now ready to be analyzed with our
approach. First, we identify the chunk containing the buggy
line(s). We shall discuss later the scenario where multiple
chunks are corrected in a single run. Next we identify the first
occurrence of this chunk in the logged record, and discard
everything after this occurrence. This is necessary because 
once a buggy chunk is visited,
all subsequent steps are unreliable.

Next, we extract the statistics $m,k$ 
 and $n_i$'s from
the accumulated records, and use these to compute and maximize $\ell(p)$
for $p\in(0,1).$

It should be noted that all the steps can be easily automated,
and hardly cause any disruption to the usual work flow
of the programmer in charge of debugging. This seamless
integration with the existing habits of programmers is a major
strong point of {\bf RESID}.

\section{Results}
The proper evaluation of {\bf RESID} may only be done in an
industrial set up where a large, complex software is actually
being debugged. In this paper we present the results using a
simulated 
toy example. 
We start with a C program, and simulate a bug in each chunk with
probability $p\alpha^r$ where $p$ is some chosen value
of the parameter of interest, 
$\alpha=0.9$ (chosen arbitrarily) measures debugging inefficiency,
and $r$ is the number of times this chunk has already been
debugged. 

The first example
is a simple program consisting of just 4 chunks as shown in the
flowchart in {\rm Fig \ref{im:flow2.eps}}. Each rhombus represents an
{\tt if}-clause, and the rounded rectangle represents a loop that
extends up to the circular connector. In the simulated runs we
take each branch in an {\tt if}-clause with equal
probability. Also the loop is run for a random number of steps
generated uniformly from $\{1,...,100\}.$

\chhobi{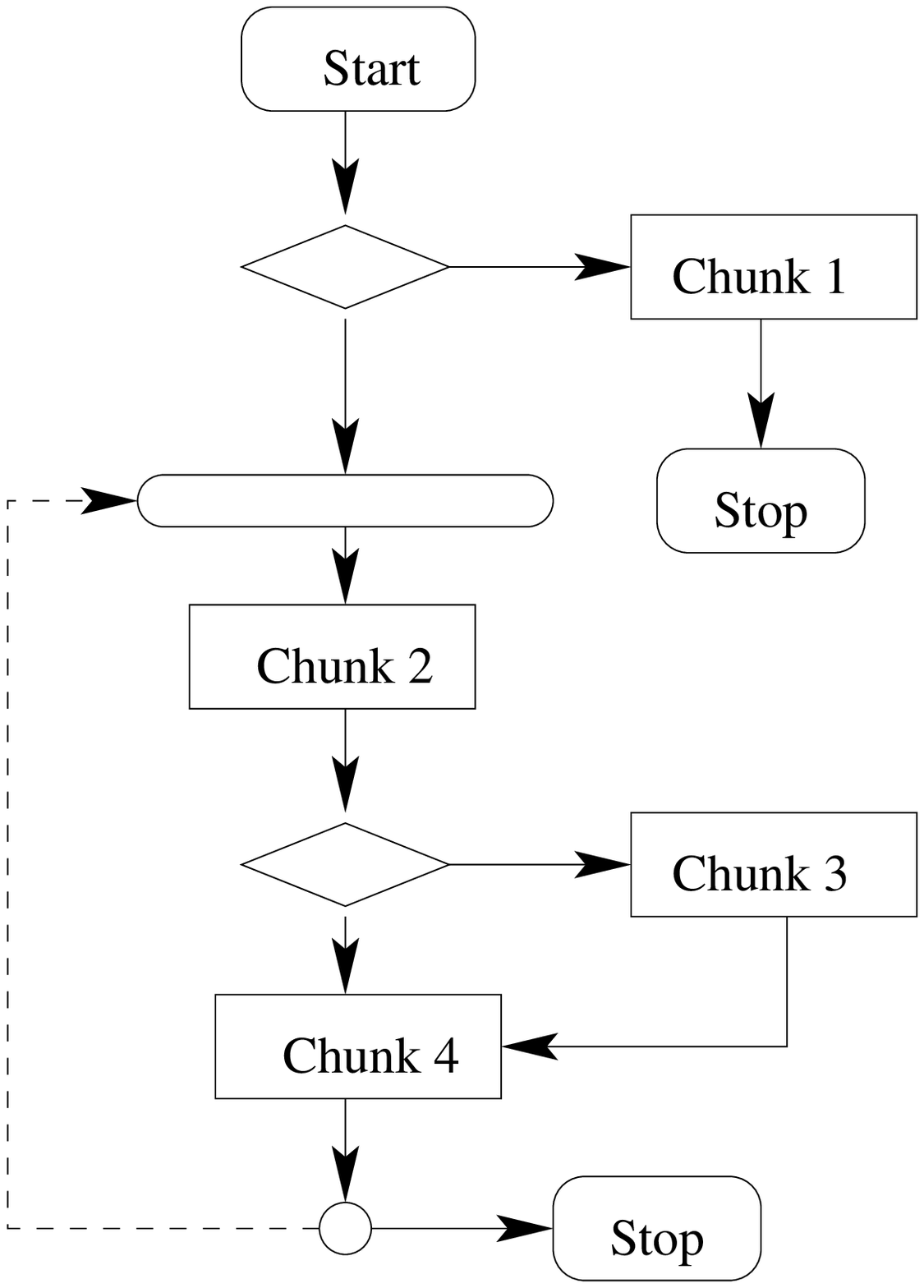}
       {0.4}
{0}
       {A simple flowchart}

The debugging session is run 100 times each with the
values $p=0.2,0.4,0.6$ and $0.8.$ 
The log-likelihood functions are shown
in {\rm Fig \ref{im:llik.eps}}.

\chhobi{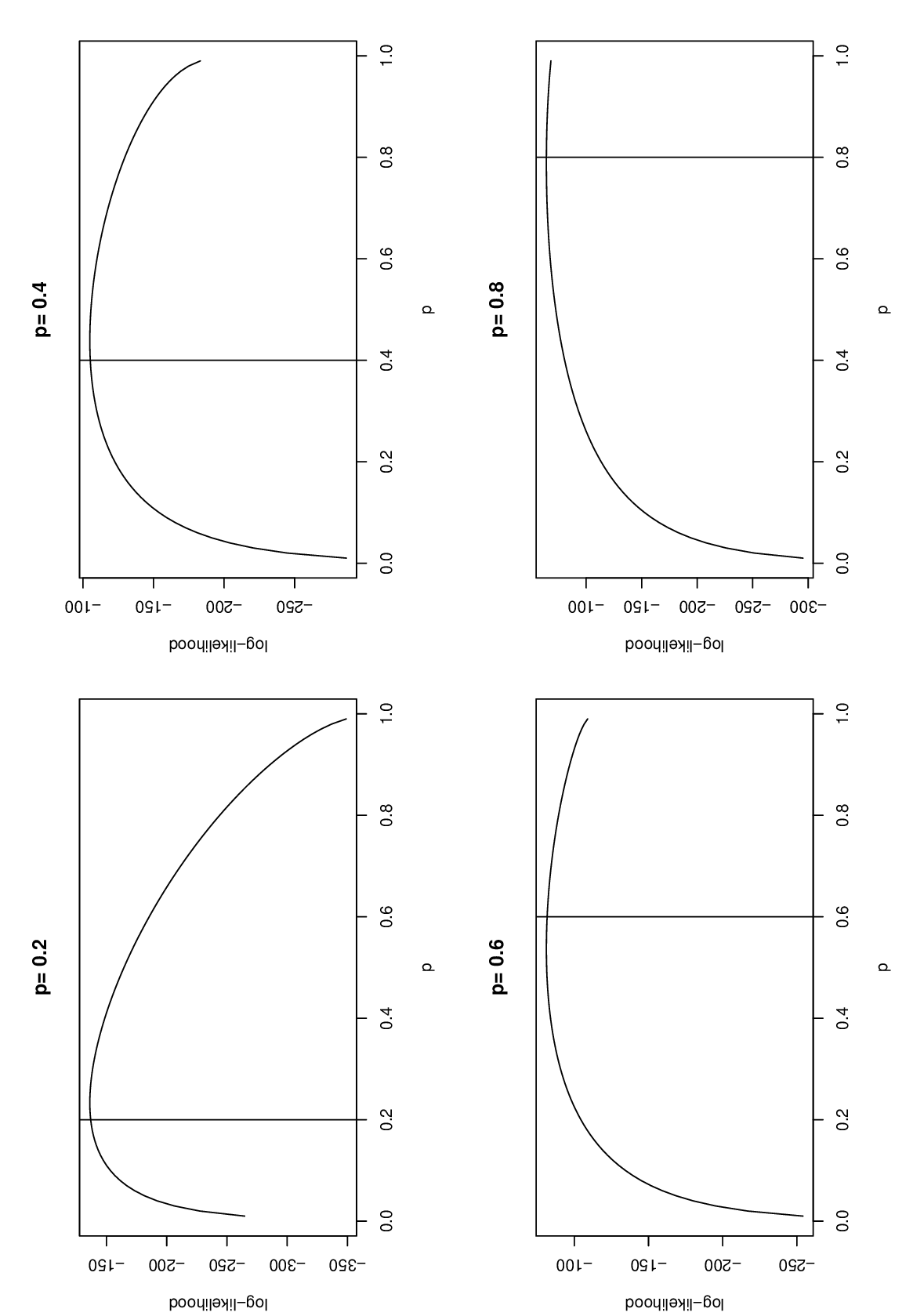}
       {0.4}
{-90}
       {Simulated log-likelihood functions}
 
Each time the peak is very near the actual value of $p$
(shown with a vertical line).

In order to estimate the variances and also to judge the effect
of $\alpha$ we apply 50-run simulations  100 times for three different
values $p=0.3,0.6,0.9$ and three different
values of $\alpha=0.3,0.6,0.9.$ The resulting mean MLE's are as follows.

\begin{center}

\begin{tabular}{c|ccccccccccccccc}
\hline\hline

&&$\alpha=0.3$ & $\alpha=0.3$ & $\alpha=0.3$\\
\hline
$p=0.3$ & 0.3269 & 0.3161 & 0.3057\\
$p=0.6$ & 0.6096 & 0.6080 & 0.6155\\
$p=0.9$ & 0.9178 & 0.9083 & 0.9006
\\
\hline
\end{tabular}
\end{center}

The corresponding variances are 

\begin{center}

\begin{tabular}{c|ccccccccccccccc}
\hline\hline

&&$\alpha=0.3$ & $\alpha=0.3$ & $\alpha=0.3$\\
\hline
$p=0.3$ & 0.005 & 0.003 & 0.002\\
$p=0.6$ & 0.015 & 0.010 & 0.006\\
$p=0.9$ & 0.009 & 0.008 & 0.006
\\
\hline
\end{tabular}
\end{center}

\section{Variants}

The main aim of this paper is to present practically applicable
software debugging procedure. In view of practical implementation
the main idea presented in the last section may need to be
changed to some extent. We discuss some such variations in this
section. 
\subsection{Chunk-specific bug probabilities}
The assumption that each chunk has the same {\em a priori}
chance of being buggy may be stretching imagination a bit too
far. After all, a chunk consisting of just a few lines of
initialization is lot less likely to spring a bug than a chunk
containing many lines of complex code. Yet allowing each chunk
to have its own $p$ will cause an explosion in the number of
parameters. A reasonable balance may be obtained by positing that
that the chance of bugginess of a chunk is related to the number
of lines in it as follows:
\begin{equation*}\tag{{\bf *}}\label{*}
P(\text{a chunk is buggy}) = 1-(1-p)^K,
\end{equation*}
where $K$ is the number of lines in the chunk,
and $p\in(0,1).$ This formula may be motivated by
considering each line to have probability $p$ of springing a
bug, and assuming that the event that one line is buggy is
independent of another line being buggy. Then \eqref{*} gives
the chance of the chunk containing at least a single bug. 

In this case the likelihood function is of the form
$$
L(p) \propto p^m \prod_{i=0}^k \prod_{j=1}^{n_i} 
(1-(1-(1-p)^{K_{ij}})\alpha^i).
$$
As before this is just the form. Trying to interpret the
quantities like $K_{ij}$ would not lead to any easy way to
compute it for a given data set. One must rely on automated tools
for its evaluation. Complicated as it is, the log-likelihood
function still obeys the lemmas given earlier.

\subsection{Classification of chunks}
The inefficiency of debugging may depend on the type of code a chunk
contains. A chunk involving numerical analysis is
typical harder to debug than one consisting of some
initialization commands. Accordingly it may be possible to
classify all the chunks into a small number of broad categories ,
and assign different debugging inefficiency factors to these.

\subsection{Multi-chunk debugging}
After a program terminates in an undesirable fashion ({\em e.g.},
crashes or gives wrong output), a programmer has to carefully go
through the code to identify the corresponding bug. In some
rare situations the bug may not be confined in a single
chunk, as  we have assumed so far. But our approach can be easily
adapted to such a scenario. The programmer is to just record the
lines that needed correction. All the corresponding chunks are
then marked as buggy, and the logged record for the run is
truncated up to and including the first occurrence of {\em any}
of these chunks. Also the chance of these chunks still remaining
buggy is updated by scaling down the current probabilities with a
factor $\alpha.$

The corresponding change to the log-likelihood is notationally
cumbersome, but easily effected in a computer.

\subsection{Bug detected, but not removed}
Since software debugging is often done under resource
constraints, not all bugs whose presence can be detected can be
removed. This is especially true about bugs that are deemed
more ``esoteric'' in nature. In such a case we can still apply
our approach, but we do not scale down the probability for the
chunk.  

\section{Conclusion}
In this paper we have proposed a new software reliability
technique. The technique seeks to integrate {\em ad
  hoc} practical debugging methods with statistical
modelling. We have mentioned how a classical debugging session
can be easily cast into the new approach with the use of the
software tools. 

The author likes to thank Prof Anup Dewanji for introducing him
to this area of research. 
 
\section{Appendix}
Throughout the paper we have made the assumption that the
existence of 
bug(s) in  a chunk/line  is independent of the existence of a bug
in another chunk/line. We shall try to justify this heuristic
statement here in terms of rigorous probability argument. 
Notice that when we say that "a chunk is found buggy" we actually mean
two things: 

\begin{enumerate}

\item 
that the chunk contains a bug,

\item 
that this bug is triggered by the data we are using.

\end{enumerate}

\newcommand{\calX}{{\mathcal X}}
\newcommand{\calY}{{\mathcal Y}}
\newcommand{\calF}{{\mathcal F}}
\newcommand{\calXY}{\calX\otimes\calY}
 So there are two
sources of randomness. The former is introduced by the
 programmer, the latter by the user. We shall accordingly employ
 two probability spaces $({\calX},{\calF}_x,P_x)$
 and $({\calY},{\calF}_y,P_y)$ for the programmer
 and user, respectively. We may think of  $\calX$ as all
ways of creating bugs available to the
programmer. Similarly, $\calY$ denotes the set of all
possible user inputs. We make the following assumption.

{\bf Assumptions:} 

\begin{enumerate}

\item The user and the programmer behave
independently.
\item Let $A_1,A_2\subseteq\calX$ be the events that there are
  bugs in two (disjoint) chunks. Then $A_1,A_2$ are mutually
  independent.
\item Let $B_1,B_2\subseteq\calY$ be the events that the user
  chooses an input that triggers any two distinct
  bugs. Then $B_1,B_2$ are independent.

\end{enumerate}

Thanks to the first assumption we are working in the product
space $(\calXY,\calF_x\otimes\calF_y,P_{xy}\equiv P_x\otimes P_y).$

If a user encounters a bug in a particular chunk, this event is $$
A\times B \subseteq \calXY,
$$ 
where $A\subseteq\calX$ is the event that the programmer has
left a bug in that chunk, and $B\subseteq\calY$ denotes the
event that the user happens to have chosen an input to trigger
it.

Thus when we say that the chance of finding a given chunk to be
buggy is $p$ we actually mean $P_{xy}(A\times B) = p,$
and {\em not} $P_x(A) = p.$

Now let us fix any two distinct chunks and accordingly
define $A_i\in\calF_x$ as the event of chunk $i$
containing 
a bug. Also let $B_i\in\calF_y$ be the event that the
user chooses an input value that detects a bug (if any) in chunk $i.$

Let $C_i\in\calF_{xy}$ be the event that the user
actually encounters a bug in chunk $i.$ Then  $C_1,C_2$ are independent in the product
space, because
\begin{eqnarray*}
P_{xy}(C_1\cap C_2) 
&=& P_{xy}((A_1\times B_1)\cap (A_2\times B_2))\\
&=& P_{xy}((A_1\cap A_2)\times (B_1\cap B_2))\\
&=& P_x(A_1\cap A_2)P_y(B_1\cap B_2)\\
&=& P_x(A_1)P_x(A_2)P_y(B_1)P_y(B_2)\\
&=& P_{xy}(A_1\times B_1)P_{xy}(A_2\times B_2)\\
&=& P_{xy}(C_1)P_{xy}(C_2).
\end{eqnarray*}

\bibliographystyle{plain}
\bibliography{ref}

\begin{thebibliography}{10}

\bibitem{appel}
Andrew~W. Appel and Maia Ginsburg.
\newblock {\em Modern Compiler Implementation in C}.
\newblock Cambridge University Press, New York, NY, USA, 2004.

\bibitem{bolandshort}
P.~Boland.
\newblock Challenges in software reliability and testing.
\newblock In {\em Third International Conference on Mathematical Methods in
  Reliability}, June 17--20 2002.

\bibitem{dewanji}
A.~Dewanji, T.K. Nayak, and P.K. Sen.
\newblock Estimating the number of components of a system of superimposed
  renewal processes.
\newblock {\em Sankhya, Series A}, 57(3):486--499, 1995.

\bibitem{goel}
A.L. Goel and K.~Okumoto.
\newblock Time dependent error detection rate model for software reliability
  and other performance measures.
\newblock {\em IEEE Transactions in Reliability}, R-28:206--211, 1979.

\bibitem{jelinski}
Z.~Jelinski and P.~Moranda.
\newblock Software reliability research.
\newblock In W.~Freiberger, editor, {\em Statistical Computer Performance
  Evaluation}, pages 465--84. New York: Academic, 1972.

\bibitem{moranda}
P.B. Moranda.
\newblock Prediction of software reliability software during debugging.
\newblock In {\em Proceedings on the 1975 Annual Reliability and
  Maintainability Symposium}, pages 327--32, 1975.

\bibitem{musabook}
J.D. Musa, A.~Iannino, and K.~Okumoto.
\newblock {\em Software Reliability, Measurement, Prediction, Application}.
\newblock New York: Wiley, 1987.

\bibitem{nayak}
T.K. Nayak.
\newblock Estimating population size by recapture sampling.
\newblock {\em Biometrika}, 75(1):113--20, 1988.

\bibitem{singsoyer}
N.D. Singpurwalla and R.~Soyer.
\newblock Assessing (software) reliability growth using a random coefficient
  autoregressive process and its ramifications.
\newblock {\em IEEE Transactions on Software Engineering}, 11:1456--1464, 1985.

\bibitem{singpurwalla}
N.D. Singpurwalla and S.P. Wilson.
\newblock {\em Statistical Methods in Software Engineering}.
\newblock Springer,New York, 1999.

\bibitem{xie}
M.~Xie.
\newblock Software reliability models - past, present and future.
\newblock In N.~Limnios and M.~Nikulin, editors, {\em Recent Advances in
  Reliability Theory: Methodology, Practice and Inference}, pages 323--340.
  Birkhauser, 2000.

\end{thebibliography}

\end{document}